\documentclass[preprint,showpacs,preprintnumbers,amsmath,amssymb]{revtex4}

\usepackage{graphicx}% Include figure files
\usepackage{dcolumn}% Align table columns on decimal point
\usepackage{bm}% bold math

\input epsf.tex
\def\DESepsf(#1 width #2){\epsfxsize=#2 \epsfbox{#1}}
\begin{document}

%\preprint{APS/123-QED}

\title{Evidence for Factorization in Three-body
$\overline B\to D^{(*)}\,K^-\,K^0$ Decays}
% Force line breaks with \\
\vspace{10cm}

\author{Chun-Khiang Chua, Wei-Shu Hou, Shiue-Yuan Shiau,\\ Shang-Yuu
Tsai}

%\altaffiliation[Also at ]{Physics Department, XYZ University.}

%\author{Shang-Yuu Tsai}
\affiliation{%
Department of Physics,
National Taiwan University,\\
Taipei, Taiwan 10764,
Republic of China
}%

%% \homepage{http://www.Second.institution.edu/~Charlie.Author}

\date{\today}% It is always \today, today,

  %Bottom mesons

\begin{abstract}
Motivated by recent experimental results, we use a factorization approach 
to study the three-body $\overline B\to D^{(*)}K^-K^{0}$ decay modes. 
Two mechanisms are proposed for kaon pair production: 
current-produced (from vacuum) and transition (from $B$ meson). 
The $\overline B {}^0\to D^{(*)+}K^-K^0$ decay is 
governed solely by the current-produced mechanism. 
As the kaon pair can be produced only by the vector current,
the matrix element can be extracted from 
$e^+e^-\to K\overline K$ processes via isospin relations. 
The decay rates obtained this way are in good agreement with experiment.
Both current-produced and transition processes contribute to 
$B^-\to D^{(*)0}K^- K^0$ decays. 
By using QCD counting rules and 
the measured $B^-\to D^{(*)0} K^- K^0$ decay rates, 
the measured decay spectra can be understood.
\end{abstract}

\maketitle

%%%%%%%%%%%%%%%%%%%%%%
\section{Introduction}
\label{sec:Intro}
%%%%%%%%%%%%%%%%%%%%%%

The Belle Collaboration reported recently the first observation of
$\overline B\to D^{(*)}K^-K^{(*)0}$ decays~\cite{Drutskoy:2002ib}, 
with branching fractions at the level of $10^{-4}-10^{-3}$. 
Angular analysis of the $K^-K^{(*)0}$ subsystem reveals that 
$K^-K^0$ and $K^-K^{*0}$ meson pairs are dominantly 
$J^P=1^-$ and $J^P=1^+$, respectively. 
While there is no sign of decay via resonance for the $K^-K^0$ pair, data suggest a dominant $a_1(1260)$ resonance contribution
in the production of the $K^-K^{*0}$ pair.
Both the $K^-K^{*0}$ and the $K^-K^0$ pair mass spectra of the
$D^{(*)}K^-K^{*0}$ and $D^0K^-K^0$ modes show a maximum near threshold.

These processes can be described by conventional 
$b\to c\bar ud$ diagrams with additional $s\bar s$ pair creation. 
For example, in the $\overline B {}^0\to D^{(*)+}K^- K^{(*)0}$ case,
the spectator quark ($\bar d$) ends up in the $D^{(*)+}$ meson,
while the $\bar u d$ pair ends up in the kaon pair.
It is similar to the observed three-body baryonic mode 
$B^0\to D^{*-}p\bar n$~\cite{Anderson:2000tz}, where 
the nucleon-antinucleon pair replaces the $K^-K^{0(*)}$ pair in the above. 
A generalized factorization approach~\cite{Chua:2001vh}
has been applied to study this three-body baryonic mode, 
where the amplitude is factorized into 
a current-produced $p\bar n$ part and 
a $B^0\to D^{*-}$ transition part.

A crucial ingredient in the factorization approach to 
the $B^0\to D^{*-}p\bar n$ mode is the knowledge of 
the time-like nucleon form factors. Although 
the time-like axial nucleon form factor data is still not available 
hence making the axial current contribution incalculable, 
the vector current induced nucleon form factors can be obtained 
via isospin rotation from their EM counterparts, 
where data is quite abundant.
%The data of nucleon EM form factor exhibits large momentum
%transfer suppression, which conforms to prediction by perturbative
%QCD~(PQCD).
By utilizing nucleon EM form factor data, 
it was shown that the vector current contribution 
can account for up to $60\%$ of the observed $B^0\to D^{*-}p\bar n$ rate. 
The predicted $p\bar n$ mass spectrum shows 
threshold enhancement effect as expected~\cite{Hou:2000bz}. 
This can be seen as rooted in the near-threshold behavior of
the nucleon form factors, whose appearance can be traced back to
the application of factorization and QCD counting rule,
which has been confirmed in the nucleon EM data. 
The total rate might be fully understood once 
the axial nucleon form factor becomes available. 
Alternatively, it was proposed that one can 
extract the axial nucleon form factor from 
future $B^0\to D^{*-}p\bar n$ data. 
A factorization and pole model approach has 
recently been employed to study $B\to \overline D^{(*)}p\bar n$, 
$\overline D^{(*)}p\bar p$ modes~\cite{Cheng:2002fp}, 
and some information on axial form factor is extracted.

With success in the three-body baryonic modes, and the 
encouragingly similar threshold enhancement behavior in 
the $\overline B\to D^{(*)}K^-K^{(*)0}$ modes~\cite{Drutskoy:2002ib}, 
we apply the factorization approach to study these three-body decays. 
For the $\overline B {}^0\to D^{(*)+} K^- K^0$ modes, the fact that 
the $K^- K^0$ pair is observed only in the $J^P=1^-$ state 
already supports the factorization picture, 
since {\it only} the vector current can 
produce the kaon pair under factorization.
With no axial current contribution, the amplitude can be predicted
by using kaon EM form factors through isospin relations.
This is in contrast with the $B\to D^{*-}p\bar n$ case 
where the axial current also contributes.
Thus, with no tuning of parameters, 
the $\overline B {}^0\to D^{(*)+} K^- K^0$ modes 
provide a useful testing ground of the factorization approach.

On the other hand, for the $B^-\to D^{(*)0} K^- K^0$ modes,
the kaon pair can also be produced from a $B$ meson 
by a current induced transition.
The situation is similar to $B\to p\bar p$ transitions 
in the $B\to p\bar p K$ case~\cite{Chua:2002wn}. 
Since the $B$ to kaon pair transition form factors are not known,
we use a parametrization motivated by QCD counting rules,
and determine these parameters from total decay rates. 
In other words, our approach is less predictive for these modes 
compared to $\overline B {}^0\to D^{(*)+} K^- K^0$.
However, the predicted decay spectra are closely related
to the QCD counting rules, which can be tested experimentally.

In this work we shall concentrate on 
the $\overline B\to D^{(*)}K^-K^0$ modes, while making only 
some comments on the $\overline B\to D^{(*)}K^-K^{*0}$ modes. 
The experimental data indicate that the $K^-K^{*0}$ subsystem is 
in a $1^+$ state and dominated by $a_1(1260)$ resonance,
hence originate from the axial current. 
But, unlike the vector current case for $K^- K^0$ production, 
we do not have independent information on the axial form factors,
so we have no control over these modes.
%However, we will sketch how one can utilize data of $B\to
%D^{(*)}K^-K^{*0}$ decays to extract axial form factors of $K^-K^{*0}$
%pair.
%

This paper is organized as follows: in the next section
we lay down the formalism of the factorization approach.
We show how to extract kaon form factors from EM data in Sec.~III,
and parametrize the transition form factor.
In Sec.~\ref{sec:Results} we show the results of our calculation, and in
the last section we make some discussions before conclusion is drawn.

%%%%%%%%%%%%%%%%%%%%%%%
%\section{Formalism}
%\label{sec:Formalism}
%%%%%%%%%%%%%%%%%%%%%%%
%
%-------------------------------
%\sub
\section{Factorization Formalism}

The relevant effective Hamiltonian for the 
$b\to c$ transition is
\begin{equation}
{\mathcal H_{\rm eff}}=
\frac{G_F}{\sqrt2} V_{cb} V_{ud}^*
\big[
   c_1(\mu)\mathcal O^c_1(\mu)+c_2(\mu)\mathcal O^c_2(\mu)\big], 
\label{eq:Heff}
\end{equation}
where $c_i(\mu)$ are the Wilson coefficients and $V_{cb}$
and $V_{ud}$ are the Cabibbo-Kobayashi-Maskawa~(CKM) matrix elements.
The four-quark operators ${\mathcal O}_i$ are products of two
$V-A$ currents, i.e.
${\mathcal O^c_1}=(\bar c b)_{V-A}\,(\bar d u)_{V-A}$ and
${\mathcal O^c_2}=(\bar d b)_{V-A}\,(\bar c u)_{V-A}$.

%-----------------------------------------------------------------------
\begin{figure*}[t!]
\includegraphics[width=4.5in]{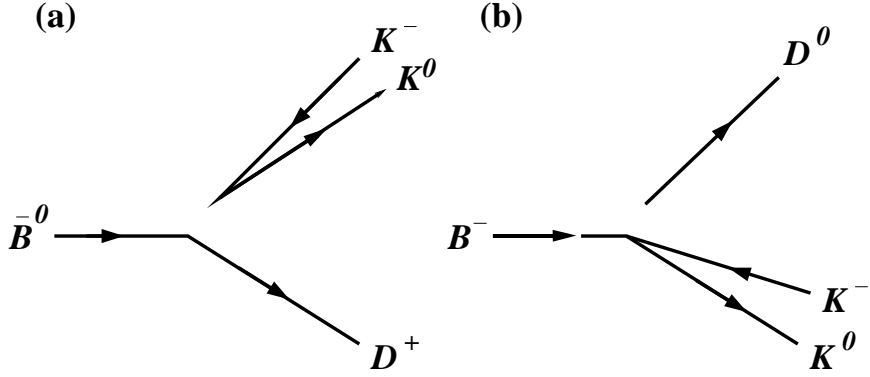} %\textwidth]{diagrams.eps}
\caption{\label{fig:diagram}
\small {(a) The current-produced
(${\mathcal J}$) and (b) the transition ($\mathcal T$)
diagrams for $\bar B^0\to D^+K^-K^0$ and $B^-\to D^0K^-K^0$ decays,
respectively.}}
\end{figure*}
%------------------------------------------------------------------------

With the factorization ansatz, the decay amplitudes for
$B\to D^{(*)}K^-K^0$ are given by
\begin{eqnarray}
{\mathcal A}(D^{(*)+}K^-K^0)&=&
\frac{G_F}{\sqrt2} V_{cb} V_{ud}^*
\,a_1 \langle D^{(*)+}|(\bar c b)_{V-A}|\overline B {}^0\rangle
      \langle K^-K^0|(\bar d u)_{V-A}|0\rangle,
\nonumber\\
{\mathcal A}(D^{(*)0}K^-K^0)&=&
\frac{G_F}{\sqrt2} V_{cb} V_{ud}^*
\big[
 a_1 \langle D^{(*)+}|(\bar c b)_{V-A}|\overline B {}^0\rangle
     \langle K^-K^0|(\bar d u)_{V-A}|0\rangle
\nonumber\\
&&\qquad\qquad
+a_2 \langle K^-K^0|(\bar d b)_{V-A}|B^-\rangle
     \langle D^{(*)0}|(\bar c u)_{V-A}|0\rangle
\big],
\label{eq:factorization}
\end{eqnarray}
where the effective coefficients are expressed as
$a_1=c_1+c_2/3$ and $a_2=c_2+c_1/3$ if naive factorization is used.
The factorized amplitudes consist of products of two matrix elements:
the case of $B$ to $D$ transition times current produced kaon pair
is called ``current-produced" (denoted as $\mathcal J$), while
the case of $B$ to kaon pair transition times current produced $D$
is called ``transition" (denoted as $\mathcal T$).
The two cases are depicted in Fig. \ref{fig:diagram}.
Note that the $\overline B {}^0\to D^{(*)+}K^-K^0$ decay
can only be current-produced, while
$B^-\to D^{(*)0} K^-K^0$ receive both contributions.
%(similar diagram for the current-produced part in the $D^{(*)0} K^-K^0$
%mode can be obtained by replacing the spectator quark 
%in Fig. \ref{fig:diagram}(a)). 

The $\langle D^{(*)}|V-A|B\rangle$ and $\langle D^{(*)}|V-A|0\rangle$
matrix elements are familiar and can be parameterized in the standard way.
We shall adopt the Bauer-Stech-Wirbel~(BSW)~\cite{BSW:physC29}
and the Melikhov-Stech~(MS)~\cite{Melikhov:2000yu} models for comparison.
The matrix elements $\langle KK|V-A|0\rangle$ 
and $\langle KK|V-A|B\rangle$ are less familiar.
They are parametrized as~\cite{BSW:physC29,Clarence:1992}
\begin{equation}
\langle K_1(p_1)K_2(p_2)\left| V^\mu\right|0\rangle=   %\bigg(
(p_1-p_2)^\mu   %-\frac{m_1^2-m_2^2}{q^2}q^\mu \bigg)
F_1^{KK}(q^2),
%+F_0^{KK}(q^2)\frac{m^2_1-m^2_2}{q^2}q^\mu,
\label{eq:Jff}
\end{equation}
\begin{eqnarray}
%\langle K_1(p_1)K_2^*(p_2,\varepsilon)\left| (V-A)_\mu\right|0\rangle&=&
%\frac{-2\,V^{KK}(q^2)}{m_1+m_2}\,
%\epsilon_{\mu\nu\alpha\beta}
%\varepsilon^{*\nu}p_1^\alpha p_2^{\beta} %,
%+i\bigg [ (m_1+m_2)\,\varepsilon_\mu^*
%A_1^{KK}(q^2)  \nonumber
%\\
%&&
%-\frac{\displaystyle {\varepsilon^*\cdot q}}{\displaystyle
%{m_1+m_2}}(p_1-p_2)_\mu %\nonumber
%A_2^{KK}(q^2)
%-2m_2\frac{\displaystyle {\varepsilon^* \cdot
%q}}{q^2}q_\mu \nonumber
%\\
%&&
%\times
%\left(A_3^{KK}(q^2)-A_0^{KK}(q^2)\right)
%\bigg ],\label{eq:ff2}
%\\
\langle K_1(p_1)K_2(p_2)|(V-A)_\mu| B(p_B)\rangle &=&
i    %\biggl[r(q^2)(p_B-p_1-p_2)_\mu+w_+(q^2)(p_2+p_1)_\mu+
w_-(q^2) (p_2-p_1)_\mu    %\biggr]
+\,h(q^2)\,\epsilon_{\mu\nu\alpha\beta}\,p_B^\nu q^\alpha
(p_2-p_1)^\beta,
\label{eq:Tff}
\end{eqnarray}
where $q\equiv p_1+p_2$, and we have 
dropped $m^2_1-m^2_2$ dependent terms in Eq.~(\ref{eq:Jff})
by assuming isospin symmetry.
%, and $K_1$, $K_2$ have been used to stand for the original kaon pair. 
Since $\langle K^- K^0|(\bar d u)_A|0\rangle = 0$ 
from Lorentz covariance and parity, 
only the vector current contributes to Eq.~(\ref{eq:Jff}),
which explains the experimental observation of $J^P = 1^-$ 
for $K^-K^0$ pair in $\overline B{}^0\to D^{(*)+} K^-K^0$. 
Data also show that the kaon pair for the $B^-\to D^{(*)0} K^-K^0$ modes
is also in a $1^-$ configuration~\cite{Drutskoy:2002ib}.
Since the transition amplitudes now also contributes,
we have used this experimental fact to drop the
$(p_B -q)_\mu$ and $q_\mu$ dependent terms in Eq.~(\ref{eq:Tff}),
as they would lead to other quantum numbers for the kaon pair.
This greatly simplifies the work.

We now show how to use data on time-like kaon EM form factors 
and an isospin relation to obtain $F^{KK}_1(q^2)$. 
We then give a simple parameterization on
$B\to KK$ transition form factors motivated by QCD counting rules.

%-------------------------------------------------------------------
\section{Current-produced and Transition $K\overline K$ Form Factors}
%-------------------------------------------------------------------

\subsection{Isospin Relation and Kaon Electromagnetic Form Factor}

The $D^{(*)+}K^-K^0$ modes contain only current-produced
amplitudes ${\mathcal J}$, as can be seen from Eq.~(\ref{eq:factorization})
and depicted in Fig.~\ref{fig:diagram}. 
Since only vector current contributes,
we can use kaon EM form factor data to obtain the weak
vector form factor via the isospin relation 
\begin{equation}
F^{KK}_1(q^2)=F_{K^+}(q^2)-F_{K^0}(q^2),
\label{eq:iso_relation}
\end{equation}
where $F_{K^+}$, $F_{K^0}$ are the EM form factors of the
charged and neutral kaons, respectively.

\begin{figure*}[t]
\centerline{\includegraphics[width=3.45in]{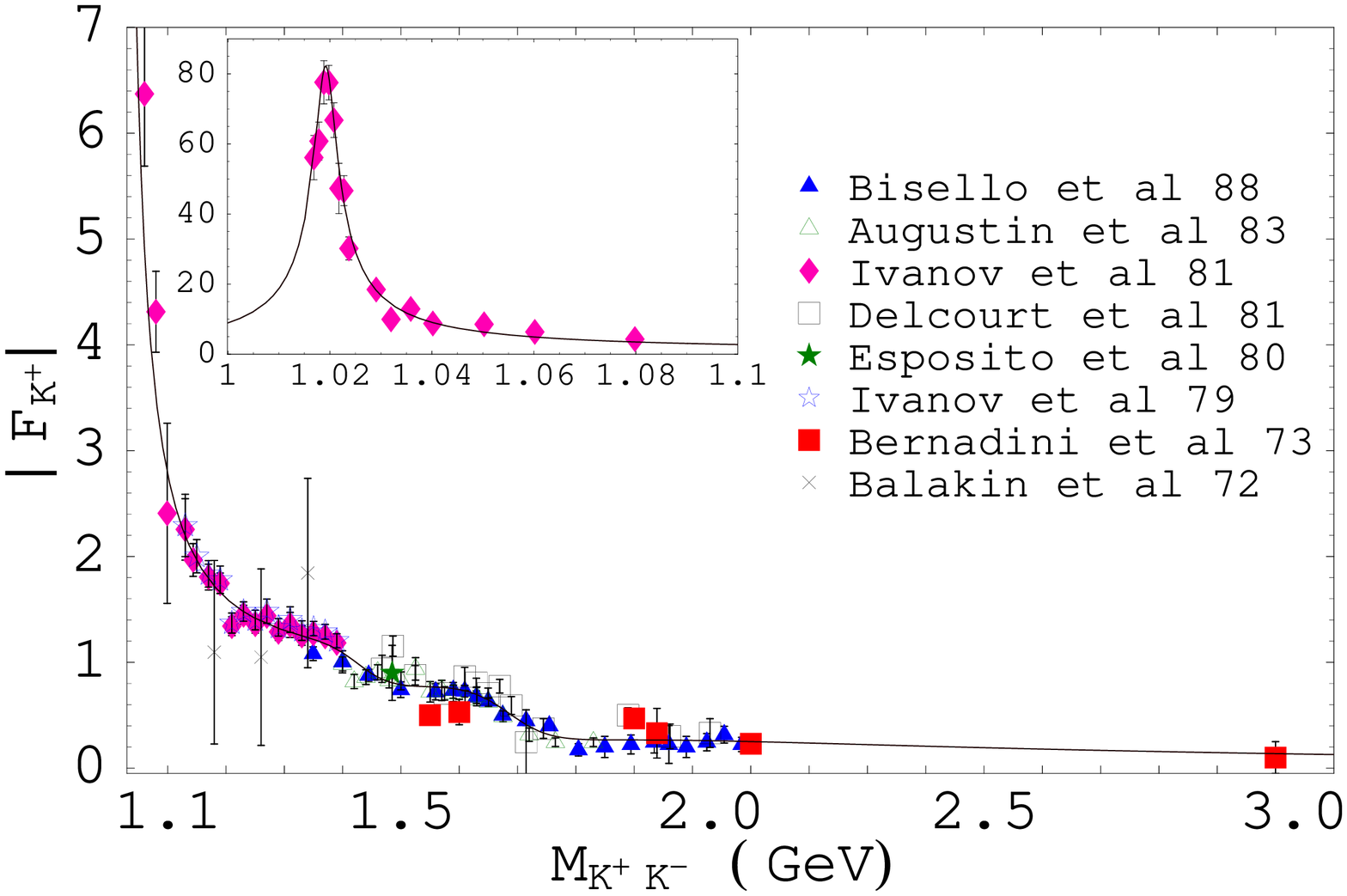} \hskip-0.8cm
\includegraphics[width=3.45in]{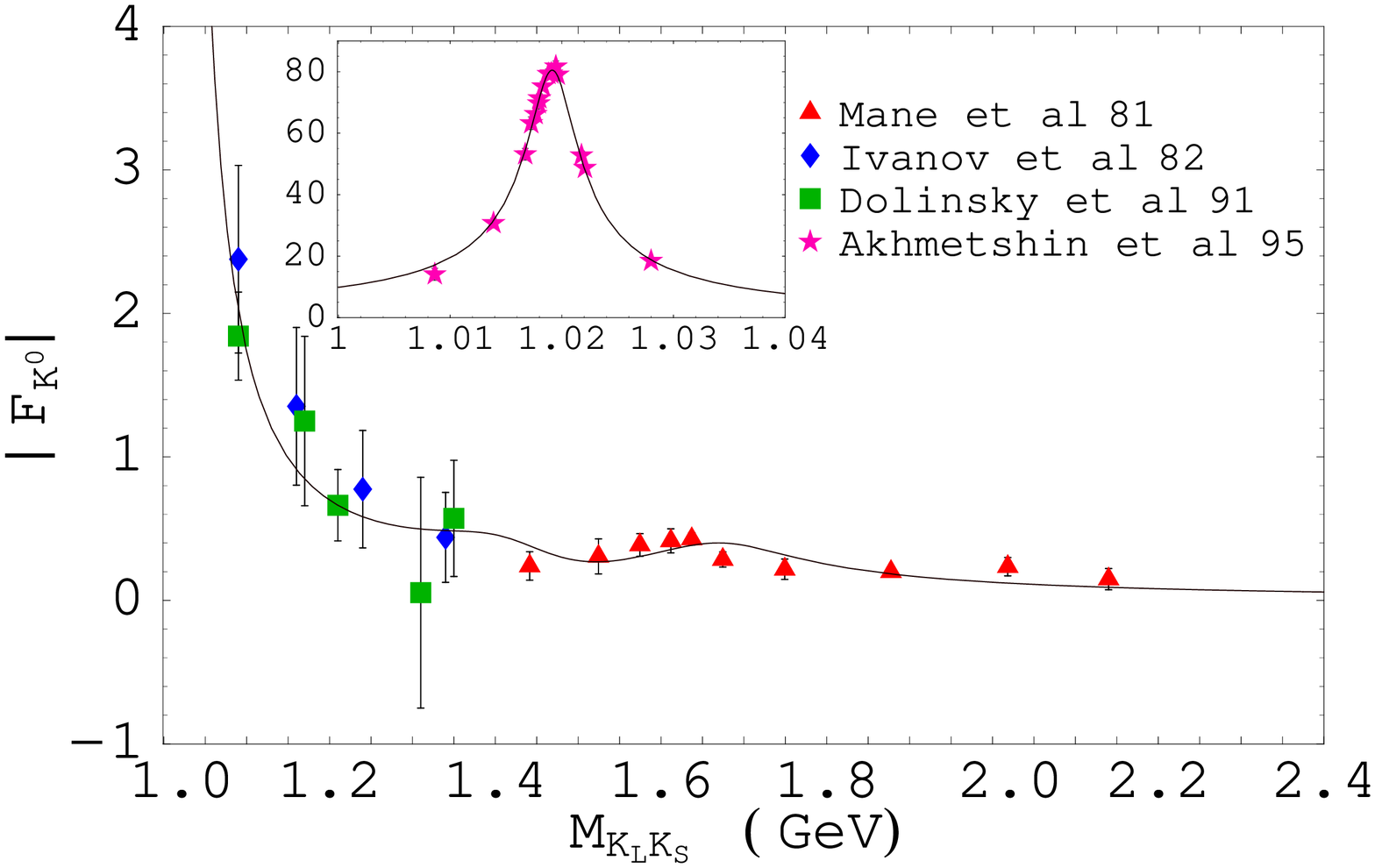}}
\caption{\label{fig:KKff}\small {
 Time-like $|F_{K^+}|$ (left) and $|F_{K^0}|$ (right) form factor data, 
 where the inset is for $\phi$ region.
 They are fitted by Eqs.~(\ref{eq:kkformf}),~(\ref{eq:k0k0formf}),
 respectively.}}
\end{figure*}

The kaon EM form factors have been measured for both space-like and time-like
regions~\cite{Balakin:vg,Bisello:1988ez,Mane:1980ep,Akhmetshin:vz},
where processes $e^+e^-\to K^+K^-$, $K_LK_S$ provide the time-like data.
The time-like $|F_{K^+}|$ and $|F_{K^0}|$ form factor data 
are given in Fig.~\ref{fig:KKff} in the energy region
$M_{K^+K^-},M_{K_L K_S}=1 \sim 3$ GeV. 
The structure is complicated in the $1\sim 2.1$ GeV range,
revealing both resonant as well as non-resonant contributions.
A sharp $\phi(1020)$ peak is shown in the insets.
The form factors drop quickly above 1.02 GeV, 
but a slower damping takes over for larger $M_{KK}$,
and one must include $\rho$, $\omega$, $\phi$ and 
their higher resonances in modeling the form factors. 
%The $\rho(770)$ has to be considered because 
%its mass is close to $K\overline K$ threshold and it has a large width.

%---------------------------------------------------------------------
%----------------------------------------------------------------------

The asymptotic behavior in $M_{KK}$ is characteristic of power-law fall off,
and seems to satisfy a stringent asymptotic constraint~\cite{Brodsky:1974vy}
from perturbative QCD~(PQCD), 
\begin{eqnarray}
F_K(t)\longrightarrow 
%-32\pi^2f_K^2/\{\beta_0\, t\,
\frac{1}{t}\left[\ln\left(\frac{t}{\Lambda^2}\right)\right]^{-1},
%\qquad (t\to \infty )
\label{eq:asymp}
\end{eqnarray}
where $t=q^2$ and $\Lambda \sim 0.3$ GeV is the QCD scale parameter.
The $1/t$ power reflects the need for a hard gluon to redistribute 
large momentum transfer.

Since our aim is just to parametrize and fit the form factor data,
we express the EM form factors by a phenomenological model that 
combines the resonant and PQCD terms
\begin{eqnarray}
F_{K^+}(t)&=&\left(\sum_{j}{c_j\over t-m_j^2+im_j\Gamma_j(t)}\right)
\sqrt{C(t)} 
+\biggl({x_1\over t} +{x_2\over
t^2}\biggr)\left[\ln\left({t\over\Lambda^2}\right)\right]^{-1},
\label{eq:kkformf}
\\
F_{K^0}(t)&=&\sum_{j}{(-)^{I_j}\,c_j \over
t-m_j^2+im_j\Gamma_j(t)}  %\nonumber\\&&
+\left({y_1\over t}\right)
\left[\ln\left({t\over\Lambda^2}\right)\right]^{-1},
\label{eq:k0k0formf}
\end{eqnarray}
where one sums over appropriate meson poles 
and $I_j$ is the isospin of the $j$th meson.
We put very few asymptotic PQCD terms because the large $t$ data is sparse.
The Coulomb factor $C(t)$~\cite{Fadin:zw}
%can be written as
%\begin{eqnarray}
%C(t)=Z{1+Z^2/4\pi^2\over 1-e^{-Z}},\qquad
%Z={\pi\alpha\over\beta},
%\end{eqnarray}
%where $\beta$ is the kaon velocity~\cite{Fadin:zw}. 
%It comes from 
accounts for soft photon exchange
that can enhance the cross section by a few percent for low $q^2$
and is needed for the fit. 
But in obtaining $F^{KK}_1$ from $F_{K^+}$ via Eq.~(\ref{eq:iso_relation}),
we divide out this effect since there is no exchange of soft photons for $K^-K^0$.

\subsection{Fitted Kaon Weak Vector Form Factor}

A major difference with our earlier studies of baryonic modes~\cite{Chua:2001vh}
is the presence of the resonance part, which turns out to be important.
To be able to account for the rich structure in the experimental data
in $m_{K^-K^0},m_{K_L,K_S}\le 2$~GeV, we take the eight vector mesons
$\rho(770)$, $\omega(782)$,
$\phi(1020)$,
$\omega(1420)$,
$\rho(1450)$,
$\omega(1650)$,
$\phi(1680)$, $\rho(1700)$,
with their masses $m_j$ kept at the experimental values~\cite{Hagiwara:pw}.
The widths $\Gamma_j(t)$ for $j=\rho,\ \omega,\ \phi$ 
are given by~\cite{Bisello:1988ez}
\begin{eqnarray}
\Gamma_\rho(t)&=&{m_\pi^3\Gamma_\rho\over s
m_\rho}\left({t-4m_\pi^2\over m_\rho^2-4m_\pi^2}
\right)^{3/2},   
\quad\quad \Gamma_\omega(t)=\Gamma_\omega,
\nonumber
\\
\Gamma_\phi(t)&=&{m_\phi^2\Gamma_\phi\over
2s}\Biggl[\left({t-4m_{K^+}^2\over m_\phi^2-4m_{K^+}^2}
\right)^{3/2}   %\nonumber\\&&
+\left({t-4m_{K^0}^2\over m_\phi^2-4m_{K^0}^2}
\right)^{3/2}\,\Biggr],
\end{eqnarray}
where $\Gamma_{\rho,\omega,\phi}$ are the full widths~\cite{Hagiwara:pw}.
For the other higher-mass vector mesons, we simply take their
experimental width values~\cite{Hagiwara:pw}.

We should stress that our phenomenological model is devised 
to account for EM form factor data in the region of our interest.
No attempt is made for deeper theoretical understanding, 
nor for data outside the $t$-region of interest. 
We therefore do not need all the $\phi$-region data 
even though more precise measurements have been obtained, 
as one would need more sophisticated treatment of $\Gamma_\phi(t)$
 (including a $\phi\to3\pi$ phase space term)~\cite{Akhmetshin:vz} 
which would complicate our parameterization. 
Actually, our neglect of such data causes no harm since $\phi$ is an isoscalar.
While it is needed to account for $F_{K^+}$ and $F_{K^0}$, 
its effect cancels in the weak form factor $F_1^{KK}(q^2)$, 
as required by isospin symmetry.

The coefficients $c_j$ are treated as
free real parameters though by some physical conditions they are
not all independent. In the VMD framework,
$c_\rho,\ c_\omega,\ c_\phi$ are expressed as
$c_\rho=g_{\rho\gamma}g_{\rho KK}$,
$c_\omega=g_{\omega\gamma}g_{\omega KK}$ and
$c_\phi=g_{\phi\gamma}g_{\phi KK}$.
Using experimental values of the electronic
widths~\cite{Hagiwara:pw} for the coupling constants $g_{V\gamma}$
and assuming SU(3) relations with ideal mixing,
namely $g_{\rho KK}={g_{\phi KK}/\sqrt2}=g_{\omega KK}$,
we have~\cite{Bisello:1988ez}
\begin{equation}
c_\rho :c_\omega :c_\phi=1:{1\over 3}:1. 
\label{eq:ccc}
\end{equation}
For other higher-mass vector mesons, $c_j$ are
free from the above constraint.

We take Eqs.~(\ref{eq:kkformf}) and (\ref{eq:k0k0formf}) to make a
phenomenological fit to the experimental data of the charged
\cite{Balakin:vg,Bisello:1988ez}
and neutral kaon form factors
\cite{Mane:1980ep,Akhmetshin:vz}.
The best fit values are obtained by finding the minimum of the $\chi^2$
function, $\chi^2\equiv\chi^2_++\chi^2_0$, where
\begin{equation}
\chi^2_{+,0}\equiv\sum_i\frac{\left[z_i-|F_{K^{+,0}}(t_i)|\right]^2}{\sigma_i^2},
\label{eq:chisquare}
\end{equation}
where $z_{i}$s are the measured absolute values of
the time-like kaon EM form factors and $\sigma_{i}$s are the error-bars. 
Note that we fit the absolute values of the kaon form factors 
since these are what can be obtained by experiment. 
Also, due to the common resonance part in both $F_{K^+}$ and $F_{K^0}$, 
we search for the minimum of $\chi^2$ as 
a combination of these two form factors.

We find the best fit values (in unit of GeV$^2$) :
\begin{equation}
\begin{array}{lll}
c_\rho=0.363,
  & c_{\rho(1450)}=7.98\times 10^{-3},\ \
  & c_{\rho(1700)}=1.71\times10^{-3},\ \
\\
c_{\omega(1420)}=-7.64\times 10^{-2},
  & c_{\omega(1650)}=-0.116,
  & c_{\phi(1680)}=-2.0\times10^{-2},
\\
\end{array}
\label{eq:cj}
\end{equation}
and
\begin{eqnarray}
x_1=3.26~{\rm GeV}^2, \qquad x_2=-5.02~{\rm GeV}^4;
 \qquad y_1=-0.47~{\rm GeV}^2. 
\label{eq:xy}
\end{eqnarray}
Note that the ratios of $c_\phi,\ c_\omega$ with $c_\rho$ are already
fixed by Eq.~(\ref{eq:ccc}),
and the best fit value for $c_\rho$ is consistent with
Ref.~\cite{Bisello:1988ez}.
As can be seen from Fig.~\ref{fig:KKff}, 
the fit is reasonable ($\chi^2/\rm n.d.f=194/130\sim 1.5$) 
for both low and high energies.

%-----------------------------------------------------------------------
\begin{figure}[t!]
\includegraphics[width=3.5in]{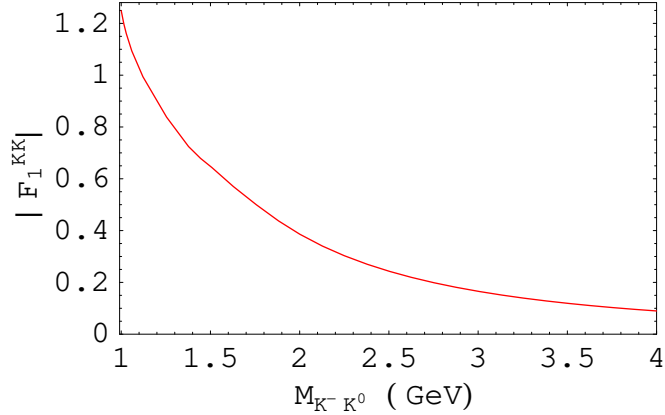}
\caption{\label{fig:KweakFF}The kaon weak vector form factor $F^{KK}_1(t)$.}
\end{figure}
%------------------------------------------------------------------------

Using Eq.~(\ref{eq:iso_relation}), we give the kaon weak vector form factor $F_1^{KK}$
in Fig.~\ref{fig:KweakFF},
with the overall factor $\sqrt{C(t)}$ in $F_{K^+}$ removed.
Contributions from poles of $I=0$ mesons cancel out,
and one is left with only the $\rho(770)$, $\rho(1450)$, $\rho(1700)$ 
and PQCD contributions.
Since $c_{\rho(1450)}$, $c_{\rho(1700)}$ are small,
$F^{KK}_1$ is modeled by the $\rho(770)$ and PQCD parts hence quite smooth.
Under factorization assumption,
this could be the reason behind the absence of structures 
in the $K^-K^0$ pair spectrum for the $\overline B {}^0\to D^{(*)+} K^- K^0$ 
decays~\cite{Drutskoy:2002ib}.

%-------------------------------------------------------------------
\subsection{\boldmath Ansatz for $B\to K^-K^0$ Transition Form Factors}
%-------------------------------------------------------------------

In addition to the current-produced matrix element,
we also need the transition matrix element
$\langle K^-K^0|(\bar d b)_{V-A}|B^-\rangle$ for
$B^-\to D^{(*)0}K^-K^0$ decay amplitudes.
Two out of four possible transition form factors are eliminated by the 
experimental observation of $J^P=1^-$ for the kaon pair~\cite{Drutskoy:2002ib}. 
There is no experimental data on the remaining form factors,
denoted $w_-(q^2)$ and $h(q^2)$ in Eq.~(\ref{eq:Tff}).

It is interesting to study 
the asymptotic behavior of these form factors from QCD counting rules.
To produce a kaon pair with large invariant mass from a decaying
$B$ meson, at least two hard gluon exchanges are needed:
one creating the $s\bar s$ pair in $K^-K^0$, 
the other kicking the spectator to catch up with 
the energetic $s$ quark to form the $K$ meson.
This gives rise to a $1/t^2$ asymptotic behavior, 
where $t$ is the $K^-K^0$ invariant mass squared. 

Resonant contributions such as from an intermediate $\rho$ pole 
should in principle be considered for the $B^-\to K^-K^0$ transition. 
This seems to be supported by the dominant $\rho$ contribution in
$F_1^{KK}(q^2)$ as we have just discussed.
However, we would need further poles such as 
$\rho(1450)$, $\rho(1700)$ to account for 
the $1/t^2$ asymptotic behavior implied by QCD counting rules.
Our experience with $F_1^{KK}$ does not help 
since these resonances are unimportant there,
while we lack other independent experimental information.
But there is as yet no clear sign of resonances in the kaon pair spectrum.
Because of this, we shall use a very simple parameterization 
solely motivated from QCD counting rules, i.e.
\begin{equation}
w_-(t)=\frac{c_{w}}{t^2},\qquad
h(t)=\frac{c_{h}}{t^2},
\label{eq:transitionFF}
\end{equation}
where $c_{w,h}$ are free parameters to be fitted by data. 

%Because of the form of Eq.~(\ref{eq:transitionFF}), the value of
%the form factors can rise unlimitedly as one approaches $t=0$
%from larger $t$, consequently the decay rate must have peak near
%threshold, which can be attributed to the application of PQCD
%counting rule. However, one should not forget that the would-have
%existed higher order terms have been absent from the
%parameterization, had them been included, and alternating in sign,
%the threshold enhancement would be cut down to a smaller size.
%This already is seen in nucleon form factors~\cite{Chua:2001vh}.

%%%%%%%%%%%%%%%%%%%%%%%%%%%%%%%%%%%%
\section{Results}\label{sec:Results}
%%%%%%%%%%%%%%%%%%%%%%%%%%%%%%%%%%%%

We use the central values of the effective coefficients 
$a_1^{\rm BSW}=0.91\pm0.08\pm0.07\,(0.86\pm0.21\pm0.07)$ and
$a_2^{\rm BSW}=0.56\pm0.31\,(0.47\pm0.41)$
extracted from~\cite{Cheng:1999tr} $\overline B {}^0\to D^{(*)+}\rho^-$,
and $B^-\to D^{(*)0}\rho^-$, respectively. 
Similarly, we use $a_1^{\rm MS}=0.935~(0.803)$ 
and $a_2^{\rm MS}=0.5~(0.553)$ for the MS form factor case.
The CKM matrix elements $V_{ud}$ and
$V_{cb}$ are taken to be $0.975$ and $0.039$, respectively.

It is useful to give first an outline of our results.
Under factorization, 
the theoretical input for the current-produced $D^+ K^-K^0$, 
$D^{*+}K^-K^0$ modes are all determined, such as $F_1^{KK}$ from EM data
and $a_1$ from the $\overline B {}^0\to D^+\rho^-, D^{*+}\rho^-$ rates.
The calculated rates turn out to be 
in good agreement with the experimental results~\cite{Drutskoy:2002ib}.
The decay spectra are predictions, which can be checked soon.
For the $D^{(*)0} K^-K^0$ case, since we have to 
fit the unknown transition form factors from the corresponding decay rates, 
it is not as solid as the purely current-produced case.
But it is interesting that the resulting decay spectra agree 
well with experimental results~\cite{Drutskoy:2002ib}.   
We comment on $D^{(*)}K^-K^{*0}$ modes before we end this section.

%----------------------------------------------------------------------
\subsection{\boldmath $\overline B {}^0\to D^{(*)+}K^-K^0$}
%----------------------------------------------------------------------

From Eqs. (\ref{eq:factorization}) and (\ref{eq:Jff}),
the $\overline B {}^0\to D^+K^-K^0$ decay amplitude is given by
\begin{equation}
{\mathcal A}(D^+K^-K^0)=\frac{G_F}{\sqrt2}\,V_{cb} V_{ud}^*\,
a_1(p_B+p_D)\cdot(p_2-p_1)%\nonumber
%\times
\,F^{BD}_1(q^2)\,F_1^{KK}(q^2),
\label{eq:ampD+KK}
\end{equation}
where $p_1$, $p_2$ stand for the momenta of $K^-$ and $K^0$, respectively. 
Factorization implies that the amplitude involves only the known 
weak kaon form factor $F^{KK}_1$ and the $B\to D$ form factor $F^{BD}_1$.
For $\overline B {}^0\to D^{*+}K^-K^0$, the decay amplitude is
\begin{eqnarray}
{\mathcal A}(D^{*+} K^- K^0)
&=&
\frac{G_F}{\sqrt2}V_{cb} V_{ud}^*\,
a_1F_1^{KK}(q^2)  %\nonumber\\&&\qquad\times
\bigg\{\frac{2\,V(q^2)}{m_B+m_{D^*}}\nonumber
\,\epsilon_{\mu\nu\alpha\beta}\,
(p_2-p_1)^\mu\varepsilon_{D^{*}}^{*\nu}p_B^\alpha
p_{D^*}^{\beta}\nonumber
\\
&&  
+\,i\,\bigg[
(m_B+m_{D^*})A_1(q^2)\,\varepsilon_{D^*}^*\cdot
(p_2-p_1)   
-\frac{A_2(q^2)\,\varepsilon_{D^*}^*\cdot
q}{m_B+m_{D^*}}\nonumber
\\
&&
\times\,(p_B+p_{D^*})
\cdot(p_2-p_1)\bigg] \bigg\},
\end{eqnarray}
which involves the $B\to D^*$ transition form factors $V$, $A_{1,2}$.
There is no tunable parameters for these two modes, hence
they provide a good test for the factorization approach.
The $p_2-p_1$ factor implies that the $K^-K^0$ pair 
must be in a P-wave state, which is 
just what the Belle experiment observes~\cite{Drutskoy:2002ib}.

%%%%%%%%%%%%%%%%%%%%%%%%%%%%%%%%%%%%%%%%%%%%%%%%%%%%%%%%%%%%%%%%%%%%%%%%
\begin{table}[t]
\caption{\label{tab:BRD+KK} ${\mathcal B}(B^-\to D^{(*)+}K^-K^0)$
 in units of $10^{-4}$, where the upper and lower limits come from scanning 
 the $\chi^2_{\rm min}+1$ region for the kaon weak form factor $F^{KK}_1$.
 }
\begin{ruledtabular}
\begin{tabular}{lccc}
 & MS &  BSW &  Experiment~\cite{Drutskoy:2002ib}
\\
\hline
$ \overline B {}^0\to D^+K^-K^0$  & $1.67^{+0.24}_{-0.21}$ 
&$1.54^{+0.22}_{-0.20}$
& $1.6\pm0.8\pm0.3$
\\
 & & & $<3.1$~($90\%$~CL)
\\
\hline
%$B^-\to D^0 K^-K^0$  & $5.5$ &$ 5.5$&$5.5\pm1.4\pm0.8$
%\\
%\hline
$\overline B {}^0\to D^{*+}K^-K^0$ 
&$2.8^{+0.30}_{-0.36}$&$3.05^{+0.32}_{-0.39}$&
$2.0\pm1.5\pm0.4$
\\
 & & & $<4.7$~($90\%$~CL)
\\
%\hline
%$B^-\to D^{*0}K^-K^0$ &$1.27$& $ 1.45$& $ 6.0\pm2.6\pm1.3$
%\\
%& & &$<11.4 ~(90\% ~CL)$
%\hline
%$M_V$~(GeV) & \multicolumn{2}{c}{5.42} & \multicolumn{2}{c}{5.32}
%\\
\end{tabular}
\end{ruledtabular}
\end{table}
%%%%%%%%%%%%%%%%%%%%%%%%%%%%%%%%%%%%%%%%%%%%%%%%%%%%%%%%%%%%%%%%%%%%%%%%

We show in Table~\ref{tab:BRD+KK} the calculated rates of these two modes.
Both turn out to be at the $10^{-4}$ level. The upper and
lower limits are from scanning the $\chi^2_{\rm min}+1$ region
for the maximum and minimum of the kaon weak form factor
$F^{KK}_1$. 
We find good agreement between factorization and experimental results, 
especially for the $D^+K^-K^0$ case.
This provides {\it evidence} that the factorization approach works.

%%%%%%%%%%%%%%%%%%%%%%%%%%%%%%%%%%%%%%%%%%%%%%%%%%%%%%%%%%%%%%%%%%%%%%%%
%\begin{table}[t]
%\caption{\label{tab:ResVSnonRes}Resonant v.s. non-resonant contributions
%in the branching fractions~(in units of $10^{-4}$) of the $\overline B 
%{}^0\to
%D^{(*)+}K^-K^0$ modes. The MS results are used for illustration.} 
%\begin{ruledtabular}
%\begin{tabular}{lccc}
%%
% & ${\mathcal B}^{\rm MS}\,$(resonant) & ${\mathcal B}^{\rm
%MS}\,$(non-resonant) & ${\mathcal B}^{\rm MS}\,$(both) \\
%\hline
%$\overline B {}^0\to D^+K^-K^0$ & $0.81$ & $0.21$ & $1.67$ \\
%$\overline B {}^0\to D^{*+}K^-K^0$ & $1.19$ & $0.41$ & $2.8$ \\
%%
%\end{tabular}
%\end{ruledtabular}
%\end{table}
%%%%%%%%%%%%%%%%%%%%%%%%%%%%%%%%%%%%%%%%%%%%%%%%%%%%%%%%%%%%%%%%%%%%%%%%

It is instructive to understand the contribution to the
$D^{(*)+}K^-K^0$ rates from the resonant and non-resonant parts of $F^{KK}_1$,
where the latter refers to the $x_{1,2}$ and $y_1$ terms in 
Eqs.~(\ref{eq:kkformf}) and (\ref{eq:k0k0formf}).
As previously noted, the $\rho$ contribution dominates the resonant part.
We find that 
the resonant part contributes 40\% (43\%) of the $D^{(*)+}K^-K^0$ rate, 
while the non-resonant part contributes 13\% (15\%).
Constructive interference between the two is needed 
to give rates close to experimental results,
hence is indeed important.

%-------------------------------------------------------------------------
\begin{figure}[t]
\includegraphics[width=3.5in]{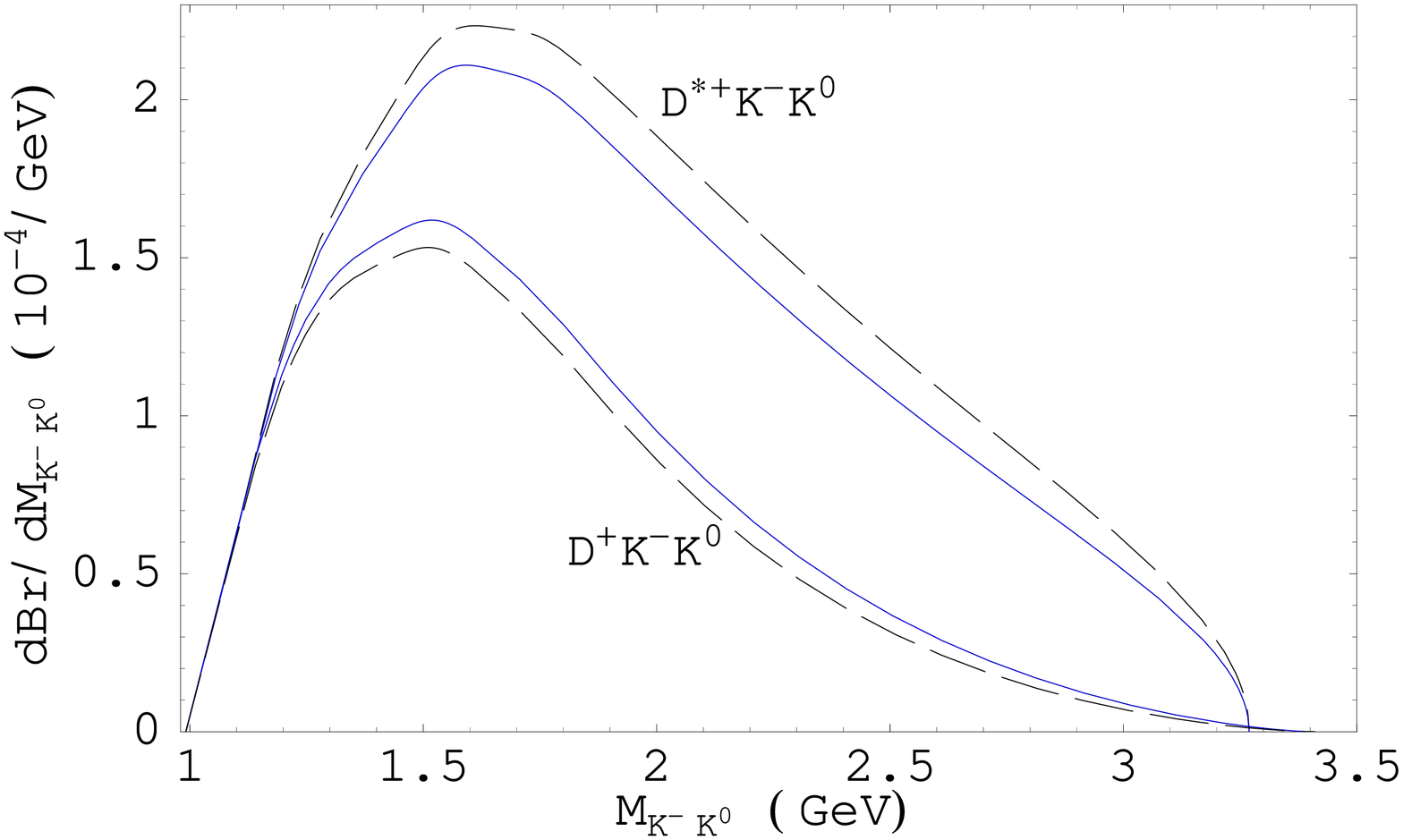}
\caption{\label{fig:D+KK} The $K^-K^0$ mass spectrum for 
$\overline B {}^0\to D^+K^-K^0$ (lower) and $D^{*+}K^-K^0$ (upper),
where solid~(dashed) line stands for using the MS~(BSW)
hadronic form factors.}
\end{figure}
%--------------------------------------------------------------------------

The $K^-K^0$ mass spectra of the $D^{(*)+}K^-K^0$ modes are 
shown in Fig.~\ref{fig:D+KK}. Both peak close to threshold, 
which is due to the near-threshold behavior of the $F^{KK}_1$ form factor 
(see Fig.~\ref{fig:KweakFF}).
There is no other clear structure, 
other than the $B\to D^*$ form factor effect at larger $q^2$.
Because of lower $D^{(*)+}$ reconstruction efficiencies,
the spectra has yet to be measured experimentally~\cite{Drutskoy:2002ib},
but our predicted spectrum can be checked soon with more data.

%-------------------------------------------------------------------------
\subsection{$ B^-\to D^{(*)0}K^-K^0$}
%-------------------------------------------------------------------------

From Eqs. (\ref{eq:factorization}), (\ref{eq:Jff}) and (\ref{eq:Tff}), 
we have the amplitudes
\begin{eqnarray}
&&{\mathcal A}(D^0K^-K^0)
=\frac{G_F}{\sqrt2}V_{cb} V_{ud}^*
\bigg[a_1\,(p_B+p_D)\cdot(p_2-p_1)  
F^{BD}_1(q^2)F_1^{KK}(q^2) 
\nonumber\\
&&\qquad\qquad\qquad\qquad\qquad\qquad
-\,a_2\,f_D\,w_{-}(q^2)p_B\cdot(p_2-p_1)\bigg],
\label{eq:ampD0KK}
\\
&&{\mathcal A}(D^{*0}K^-K^0)
=\frac{G_F}{\sqrt2} V_{cb}V_{ud}^*\Bigg\{a_1\,F_1^{KK}(q^2)
\bigg[\frac{2\,V(q^2)}{m_B+m_{D^*}}\,\epsilon_{\mu\nu\alpha\beta}\,
\varepsilon_{D^*}^{*\nu}p_B^\alpha
p_{D^*}^{\beta}(p_2-p_1)^\mu  \nonumber
\\
\nonumber\\
&&\qquad\qquad\qquad\qquad\qquad\qquad
+i\biggl(A_1(q^2)(m_B+m_{D^*})\varepsilon_{D^*}^*\cdot
(p_2-p_1)   
\nonumber\\
&&\qquad\qquad\qquad\qquad\qquad\qquad
-\frac{A_2(q^2)}{m_B+m_D^{*}}\,\varepsilon_{D^*}^*\cdot
q\,(p_B+p_{D^*})\cdot(p_2-p_1)\biggr) \bigg] \nonumber
\\
&&\qquad\qquad\qquad\qquad\qquad\qquad
+a_2f_D\,m_{D^*}\bigg(i\,w_-(q^2)\,\varepsilon_{D^*}^*\cdot
(p_2-p_1) 
\nonumber\\
&&\qquad\qquad\qquad\qquad\qquad\qquad
+h(q^2)\,\epsilon_{\mu\nu\alpha\beta}\,\varepsilon_{D^*}^{*\mu}p_B^\nu
(p_2+p_1)^\alpha(p_2-p_1)^\beta \bigg)\Bigg\}.
\label{eq:ampD*0KK}
\end{eqnarray}
%\end{widetext}
%
The $a_1$ and $a_2$ terms correspond to 
current-produced and transition parts, respectively.
There are only two form factors $w_-(q^2)$ and $h(q^2)$ 
in the transition matrix element because $K^-K^0$ is seen only in $1^-$ state.
For the $D^0 K^-K^0$ case, only $w_-(q^2)$ contributes.

Eq.~(\ref{eq:ampD0KK}) involves only one free parameter $c_{w}$
in $w_-(t)=c_{w}/t^2$, which can be obtained by fitting 
the central value of the observed rate
${\mathcal B}(D^0K^-K^0)=(5.5\pm1.4\pm0.8)\times10^{-4}$~\cite{Drutskoy:2002ib}.
We find $c_{w}^{\rm MS(BSW)}=-35.4~(-33.0)$~GeV$^3$ and $109.2~(97.4)$~GeV$^3$,
depending on constructive or destructive interference between 
the current-produced and transition amplitudes, respectively. 
If we take $h=0$ for now, Eq.~(\ref{eq:ampD*0KK}) would give
${\mathcal B}^{\rm MS(BSW)}=1.27~(1.45)\times 10^{-4}$ and 
$35.62~(24.97)\times 10^{-4}$. The latter result seems 
too large compared with the experimental result of
${\mathcal B}(D^{*0}K^-K^0)=(5.2\pm2.7\pm1.2)\times 10^{-4}$, 
hence disfavor the destructive $F_1^{KK}$--$\omega_-$ interference case
unless a fine-tuned $h(t)$ term is used.
To obtain the central value of experimental $D^{*0}K^-K^0$ rate, 
we find $c_h^{\rm MS(BSW)}= 11.3~(13.1)$~GeV$^3$ 
or $-16.1~(-18.5)$~GeV$^3$.
We summarize in Table~\ref{tab:BRD0KK} the relevant parameters and the
measured rates by experiment.

%----------------------------------------------------------------------------
\begin{table*}[t]
\caption{\label{tab:BRD0KK} 
Fitted values of transition form factor parameters $c_{w_-}$ and $c_h$,
in units of GeV$^3$, by using the central values of $D^{(*)0}K^-K^0$ rates.
}
\begin{ruledtabular}
\begin{tabular}{lccc}
 & $c_{w_-}^{\rm MS(BSW)}$ &  $c_h^{\rm MS(BSW)}$ 
 & ${\mathcal B}(10^{-4})$~\cite{Drutskoy:2002ib} \\
\hline
$B^-\to D^0 K^-K^0$  & $-35.4~(-33.0)$ & --- & $5.5\pm1.4\pm0.8$ \\
%\hline
$B^-\to D^{*0}K^-K^0$ & $-35.4~(-33.6)$ & $11.3~(13.1)$ or $-16.1~(-18.5)$
& $5.2\pm2.7\pm1.2$%~($<10.6$ at $90\%$ CL)  
\\
\end{tabular}
\end{ruledtabular}
\end{table*}
%-----------------------------------------------------------------------------

%----------------------------
\begin{figure}[b]
\includegraphics[width=4in]{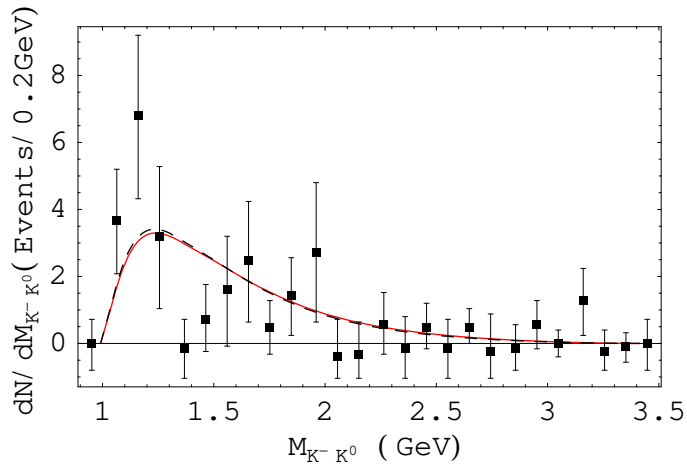}
\caption{\label{fig:d0KKdata} 
$B^-\to D^{0}K^-K^0$ spectrum, where solid (dashed) line is 
for the MS~(BSW) model, and the data is from Ref.~\cite{Drutskoy:2002ib}.}
\end{figure}
%---------------------------------------------------------------------

Both $D^0K^-K^0$ and $D^{*0}K^-K^0$ rates are used as input
and hence are not predictions.
However, we can still make predictions on the decay spectra.
By taking the fitted values of $c_{w}^{\rm MS(BSW)}$, we plot the
differential decay rate for the $B^-\to D^{0}K^-K^0$ mode in
Fig.~\ref{fig:d0KKdata} and compare with the experimental data.
The agreement is good. 
We see that the data itself shows a maximum near the $K^-K^0$ threshold, 
which can be naturally explained by our model, 
where threshold enhancement is a genuine result from the form factors
in both current-produced and transition processes.
If threshold enhancement is even more pronounced, as data seem to suggest
some structure around $m_{K^-K^0}\sim 1.4$~GeV,
then perhaps the simple approximation of 
$\omega_-(t)$, $h(t) \propto 1/t^2$ has to be reexamined.

\begin{figure}[t]
\includegraphics[width=3.5in]{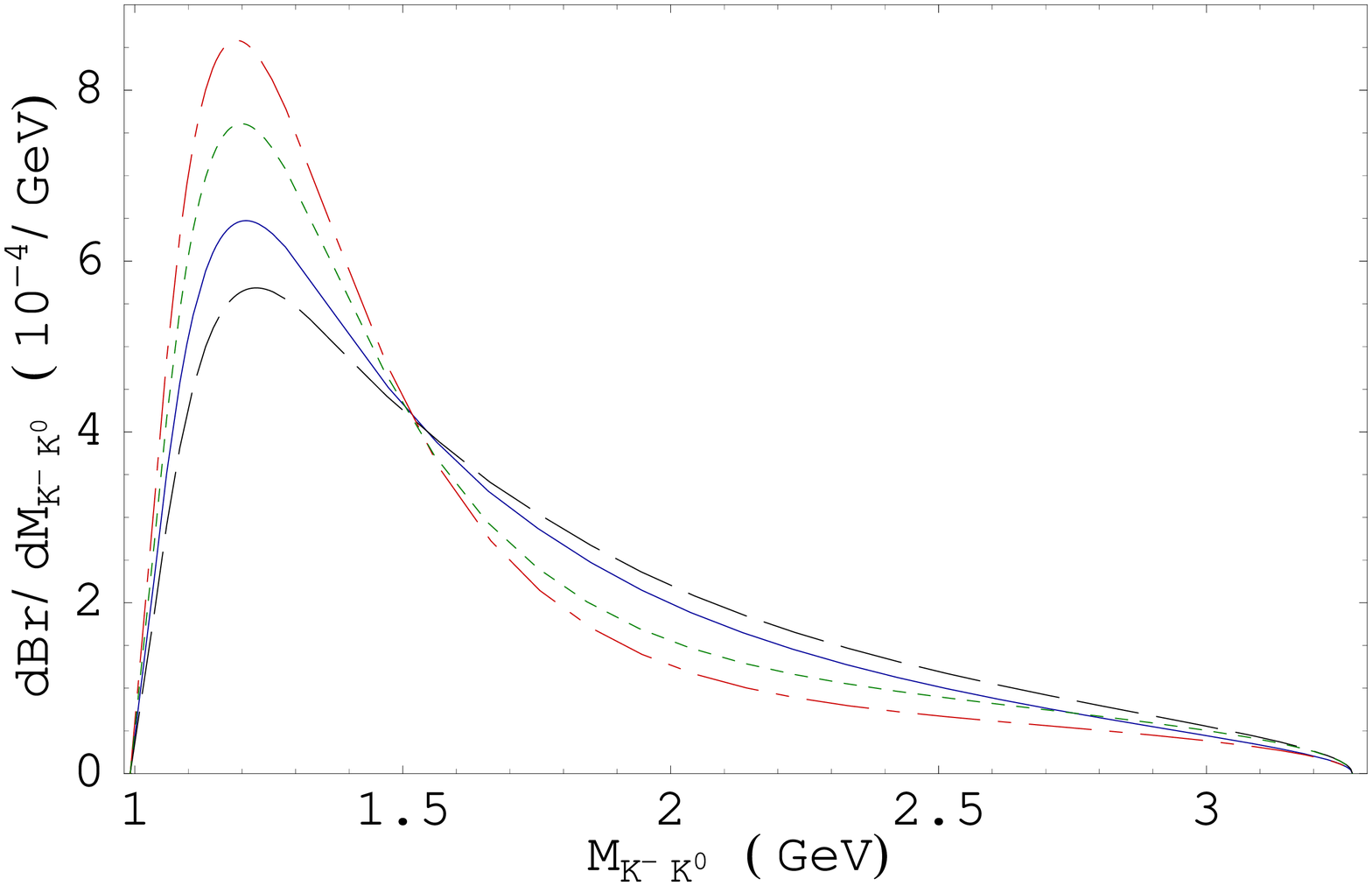}
\caption{\label{fig:charmbr2} \small {The $K^-K^0$ mass
spectrum for $\overline B {}^0\to D^{*0}K^-K^0$, where
solid, dot-dashed, dashed and dotted lines are for MS model
with $c_h=11.3,\ -16.1$~GeV$^3$ and BSW model with 
$c_h=13.1,\ -18.5$~GeV$^3$, respectively.}}
\end{figure}

The mass spectrum for $B^-\to D^{*0}K^-K^0$ mode is plotted in
Fig.~\ref{fig:charmbr2} with $c_h$ as given in Table~\ref{tab:BRD0KK}. This 
indicates the effect of the additional $h(t)$ term in the transition amplitude.
Like $B^-\to D^{0}K^-K^0$, one also has threshold enhancement, 
which again is a genuine result of both the kaon weak
form factor and the $B^-\to K^-K^0$ transition form factor.
We see clearly that the $w_{-}(t)$ and $h(t)$ form
factors contribute much in the low $K^-K^0$ mass range.

%------------------------------------------------------------------------
%--------------------------------------------------------------------------

%-------------------------------------------------------------------------
\subsection{\boldmath $ B\to D^{(*)}K^-K^{*0}$}
%-------------------------------------------------------------------------

Before we end this section, we comment on some features of 
$\overline B\to D^{(*)}K^-K^{*0}$ modes under the factorization picture.
Since there is no independent data such as EM form factors 
that one could use, we do not go into the details.

The Belle experiment already makes some interesting 
observations~\cite{Drutskoy:2002ib}:
(i) although $K^-K^{*0}$ could have $J^P=0^-,1^-,1^+$ quantum numbers,
the data prefers the $J^P=1^+$ case,
(ii) the rates seem to be dominated by the $a_1$ resonance,
but the fitted $a_1\to K^-K^{*0}$ rate is 2--5 times larger than the
CLEO result~\cite{Asner:1999kj}.
The data therefore suggest that the amplitude contains
both $a_1\to K^-K^{*0}$ resonance plus a non-resonant part.
The situation is quite similar to the $D^{(*)+}K^-K^0$ case, where
the kaon pair mainly comes from a $\rho$-pole plus a QCD motivated contribution.
The $\rho$ pole alone only gives about $40\%$ of the rate,
while constructive interference with the QCD part is quite important.
If the same picture holds in the $D^{(*)+}K^-K^{*0}$ case,
the discrepancy with the CLEO $a_1\to K^-K^{*0}$ rate may be resolved.

On the other hand, the $K^-K^{*0}$ pair is produced by an axial current,
and no EM data could be used as in the $K^-K^0$ case.
However, further theoretical tools, such as 
Weinberg sum rules~\cite{Weinberg:1967kj}, may be useful to
transfer information on vector current form factors to axial current form factor.
One can in turn extract axial current form factors from the 
$\overline B {}^0\to D^{(*)+}K^-K^{*0}$ data,
which may not be obtained by other means.

\section{Discussion and Conclusion}\label{sec:conclusion}
%%%%%%%%%%%%%%%%%%%%%%%%%%%%%%%%%%%%%%%%%%%%%%%%%%%%%%%%%

In this paper, we use factorization approach to study 
three-body $B\to D^{(*)}K^-K^0$ decays. 
There are two mechanisms of kaon pair production, 
namely current-produced and transition.
The $D^+K^-K^0$ and $D^{*+}K^-K^0$ modes involve 
only current-produced contributions. 
Under factorization, the kaon pair can only be 
generated through weak vector current, 
which can be related to EM current through isospin.
These modes provide good means to test factorization,
and the result is encouraging. 
The $D^+ K^-K^0$ and $D^{*+}K^-K^0$ 
rates in factorization approach are in good agreement with data. 

The $B^-\to D^{(*)0}K^-K^0$ decays also 
receive the transition contribution.
The form of these transition form factors are determined through
QCD counting rules, and we fix the parameters by using the measured
$D^0 K^-K^0$ and $D^{*0}K^-K^0$ decay rates.
The predicted mass spectra of the $B^-\to D^{(*)0}K^-K^0$ modes agree
well with data and exhibit
threshold enhancement as do the $\overline B {}^0\to D^{(*)+}K^-K^0$
cases. 
The $D^0K^-K^0$ spectrum shows a peak around
$m_{K^-K^0}\sim1.2$~GeV, while $D^+K^-K^0$ spectrum has a peak
around $m_{K^-K^0}\sim1.5$~GeV. This is due to the different
$1/m^2_{K^-K^0}$ behavior of current-produced and transition
form factors and can be understood from QCD counting rules.
To be specific, QCD counting rules require
the transition form factors to have a faster damping in $m_{K^-K^0}$
spectrum than the current-produced one.

Despite the success in describing the mass spectrum of the
$D^0K^-K^0$ mode, our treatment of the $B^-\to K^-K^0$ transition form 
factors may be oversimplified. Assuming the asymptotic 
form required by PQCD may be too strong an assumption, 
and might have over-enhanced the contribution from the near-threshold region.
More careful study on other possibilities, 
such as using pole models for transition via resonances, 
would be helpful in clarifying the underlying dynamics of 
$B^-\to D^{(*)0}K^-K^0$ transitions.

%In addition to account for the decay modes $B\to D^{(*)}K^-K^0$,
%we also pointed out the possibility to obtain the $K^-K^{*0}$
%form factors from $B\to D^{(*)}K^-K^{*0}$ decays. Since experiment
%has found $K^-K^{*0}$ to be produced in the $1^+$ state, by making
%use of this, together with the $K^-K^{*0}$ mass spectrum which
%can be obtained from experiment, one is able to figure out the
%$K^-K^{*0}$ form factors by inverting the process we have taken to
%evaluate for the $D^{(*)}K^-K^0$ modes.

It is interesting to compare with the treatment of 
$B^0\to D^{*-}p\bar n$~\cite{Chua:2001vh}, 
$\rho^-(\pi^-)p\bar n$~\cite{Chua:2002wn} and 
$B^-\to p\bar p K^-$ modes~\cite{Chua:2001xn}. 
The success in explaining the rates of
both $D^{*-}p\bar n$ and $p\bar p K^-$ modes, and the mass
spectrum of the latter, encouraged us to apply the factorization approach
further to the present cases. 
However, due to the lack of independent information on axial form factors, 
the difficulty to measure neutrons, and the presence of large number
of operators in the latter case, none of these modes provide 
a good testing ground for the factorization approach.
The present $D^{(*)}K^-K^0$ cases turn out to be a good place for such a test.
In particular, the $D^+K^-K^0$ and $D^{*+}K^-K^0$ modes are free from 
any undetermined parameters in the factorization approach.
The good agreement with data provide support for the
idea of factorization plus usage of isospin-related form factor data.
Further support comes from the agreement of $D^{0}K^-K^0$ spectra with data.

Even a hand-waving physical argument may be welcome to explain 
why a simple factorization approach works so well 
in these potentially complicated three-body decays.
Before we end this paper, we would like to offer one such argument.
The QCD counting rules constrain the kaon pair in the light pair mass region.
With a small invariant mass, the two kaons move colinearly and
energetically. This is certainly a conducive situation for the 
kaon pair to decouple from the recoil $D^{(*)}$ meson,
hence ``factorize".
This heuristic picture therefore resembles the 
$B$ decay to two meson case~\cite{Bjorken:kk}.
It should be noted that PQCD may not be the only explanation of
factorization~\cite{Ligeti:2001dk,Bauer:2002sh}.
The success of factorization in $\overline B\to D^{(*)} K^-K^0$
decays urges a serious study of the underlying mechanism.

\begin{acknowledgements}
We thank H.-C. Huang and M. E. Peskin for discussions. 
CKC would like to thank the SLAC theory group
for hospitality. 
This work is supported in part by the National Science Council of
R.O.C. under Grants NSC-90-2112-M-002-022, 
NSC-90-2811-M-002-038 and NSC-91-2811-M-0002-034,
the MOE CosPA Project, and the BCP Topical Program of NCTS.

\end{acknowledgements}

%%%%%%%%%%%%%%%%%%%%%%%%%%%

%%%%%%%%%%%%%%%%%%%%%
%%%%%%%%%%%%%%%%%%%%%


\begin{thebibliography}{99}


%\cite{Drutskoy:2002ib}
\bibitem{Drutskoy:2002ib}
A.~Drutskoy {\it et al.}  [Belle Collaboration],
%``Observation of B $\to$ D(*) K- K(*)0 decays,''
Phys.\ Lett.\ B {\bf 542}, 171 (2002)
[arXiv:hep-ex/0207041].
%%CITATION = HEP-EX 0207041;%%

\bibitem{Anderson:2000tz}
S.~Anderson {\it et al.}  [CLEO Collaboration],
%``First observation of the decays B0 $\to$ D*- p anti-p pi+ and B0 $\to$
%D*-  p anti-n,''
Phys.\ Rev.\ Lett.\  {\bf 86}, 2732 (2001) [arXiv:hep-ex/0009011].
%%CITATION = HEP-EX 0009011;%%

\bibitem{Chua:2001vh}
C.~K.~Chua, W.~S.~Hou and S.~Y.~Tsai,
%``Understanding B $\to$ D*- N anti-N and its implications,''
Phys.\ Rev.\ D {\bf 65}, 034003 (2002) [arXiv:hep-ph/0107110].
%%CITATION = HEP-PH 0107110;%%

%\cite{Hou:2000bz}
\bibitem{Hou:2000bz}
W.~S.~Hou and A.~Soni,
%``Pathways to rare baryonic B decays,''
Phys.\ Rev.\ Lett.\  {\bf 86}, 4247 (2001)
[arXiv:hep-ph/0008079].
%%CITATION = HEP-PH 0008079;%%


%\cite{Cheng:2002fp}
\bibitem{Cheng:2002fp}
H.~Y.~Cheng and K.~C.~Yang,
%``Three-body charmful baryonic B decays anti-B $\to$ D (D*) N anti-N,''
arXiv:hep-ph/0208185.
%%CITATION = HEP-PH 0208185;%%


%\cite{Chua:2002wn}
\bibitem{Chua:2002wn}
C.~K.~Chua, W.~S.~Hou and S.~Y.~Tsai,
%``Charmless three-body baryonic B decays,''
arXiv:hep-ph/0204185,
to appear in Phys.\ Rev.\ D.
%%CITATION = HEP-PH 0204185;%%


%%\cite{Abe:2002ds}
%\bibitem{Abe:2002ds}
%K.~Abe {\it et al.}  [Belle Collaboration],
%%``Observation of B+-  $\to$ p anti-p K+-,''
%Phys.\ Rev.\ Lett.\  {\bf 88}, 181803 (2002)
%[arXiv:hep-ex/0202017].
%%%CITATION = HEP-EX 0202017;%%


\bibitem{BSW:physC29}
%\cite{Wirbel:1985ji}
%\bibitem{Wirbel:1985ji}
M.~Wirbel, B.~Stech and M.~Bauer,
%``Exclusive Semileptonic Decays Of Heavy Mesons,''
Z.\ Phys.\ C {\bf 29}, 637 (1985).
%%CITATION = ZEPYA,C29,637;%%

%\cite{Melikhov:2000yu}
\bibitem{Melikhov:2000yu}
D.~Melikhov and B.~Stech,
%``Weak form factors for heavy meson decays: An update,''
Phys.\ Rev.\ D {\bf 62}, 014006 (2000)
[arXiv:hep-ph/0001113].
%%CITATION = HEP-PH 0001113;%%

\bibitem{Clarence:1992}
%\cite{Lee:ih}
%\bibitem{Lee:ih}
C.~L.~Lee, M.~Lu and M.~B.~Wise,
%``B(L4) And D(L4) Decay,''
Phys.\ Rev.\ D {\bf 46}, 5040 (1992).
%%CITATION = PHRVA,D46,5040;%%


%\bibitem{Diehl:2001}
%M. Diehl and G. Hiller, Phys.\ Lett.\ B  {\bf 517}, 125(2001)
%[arXiv:hep-ph/0105213].


%\bibitem{Cheng:1999xj}
%H.~Y.~Cheng and K.~C.~Yang,
%%``Implications of recent measurements of hadronic charmless B decays,''
%Phys.\ Rev.\ D {\bf 62}, 054029 (2000)
%[arXiv:hep-ph/9910291].
%%%CITATION = HEP-PH 9910291;%%



%\bibitem{text:group}
%See, for example, Fayyazuddin and Riazuddin, ``~A Modern
%Introduction to Particle Physics~'', World Scientific,2nd edition
%(2000).
%%---experimental results on B\to DKK,



%\bibitem{HYchen:2002may}
%H.~Y.~Cheng and K.~C.~Yang,
%% nonresonant three-body decays of D and B mesons,''
%arXiv:hep-ph/0205133.
%%---------------------

%\bibitem{masami:2000}
%Masami Nakagawa, Keiji Watanabe,
%%``kaon em form factor and QCD''
%Phys.\ Rev.\ C {\bf 61}, 055207 (2000).
%%%CITATION = PHRVA,D23,1152;%%

%%e^+e^-\to K^+K^-

\bibitem{Balakin:vg}
V.~E.~Balakin {\it et al.},
%``Measurement Of The Electron-Positron Annihilation Cross-Section Into 
%Pi+ Pi-, K+ K- Pairs At The Total Energy 1.18-1.34 Gev,''
Phys.\ Lett.\ B {\bf 41}, 205 (1972);
%%CITATION = PHLTA,B41,205;%%
%
%\bibitem{Bernardini:pd}
M.~Bernardini {\it et al.},
%``Proof Of Comparable K Pair And Pi Pair Production From Time-Like  
%Photons Of 1.5-Gev, 1.6-Gev, And 1.7-Gev And Determination Of The  K Meson
Phys.\ Lett.\ B {\bf 44}, 393 (1973);
%%CITATION = PHLTA,B44,393;%%
%
%\bibitem{Bernardini:pe}
M.~Bernardini {\it et al.},
%``The Time-Like Electromagnetic Form Factors Of The Charged  Pseudoscalar 
%Mesons From 1.44-Gev**2 To 9.0-Gev**2,''
Phys.\ Lett.\ B {\bf 46}, 261 (1973);
%%CITATION = PHLTA,B46,261;%%
%
%\bibitem{Ivanov:79}
P.~M.~Ivanov {\it et al.},
%``Measurement Of The Charged Kaon Form-Factor In The Energy Range 
%1.12-Gev To 1.4-Gev,''
preprint INP.-79-68 (1979);
%
%\bibitem{Esposito:bz}
B.~Esposito {\it et al.},
%``Measurements Of The Em Timelike Form-Factors For Kaon And Pion At 
%S**(1/2) = 1.5-Gev,''
Lett.\ Nuovo Cim.\  {\bf 28}, 337 (1980);
%%CITATION = NCLTA,28,337;%%
%
%\bibitem{Delcourt:1980eq}
B.~Delcourt {\it et al.},
%``Study Of The Reaction E+ E- $\to$ K+ K- In The Total Energy Range 
%1400-Mev To 2060-Mev,''
Phys.\ Lett.\ B {\bf 99}, 257 (1981);
%%CITATION = PHLTA,B99,257;%%
%
%\bibitem{Ivanov:wf}
P.~M.~Ivanov {\it et al.},
%``Measurement Of The Charged Kaon Form-Factor In The Energy Range 1.0-Gev To 1.4-Gev,''
Phys.\ Lett.\ B {\bf 107}, 297 (1981);
%%CITATION = PHLTA,B107,297;%%
%
%\bibitem{Augustin:1983ix}
J.~E.~Augustin {\it et al.},
%``A Study Of E+ E- Annihilation In The 1400-Mev To 2250-Mev Energy Range 
%With The Magnetic Detector Dm2 At Dci,''
LAL-83-21, {\it Contributed to Int. Europhysics Conf. on High
Energy Physics, Brighton, England, Jul 20-27, 1983}.

%\cite{Bisello:1988ez}
\bibitem{Bisello:1988ez}
D.~Bisello {\it et al.}  [DM2 Collaboration],
%``Study Of The Reaction E+ E- $\to$ K+ K- In The Energy Range 1350 <= 
%S**(1/2) <= 2400-Mev,''
Z.\ Phys.\ C {\bf 39}, 13 (1988).
%%CITATION = ZEPYA,C39,13;%%


%%for K_L K_S



\bibitem{Mane:1980ep}
F.~Mane {\it et al.},
%``Study Of The Reaction E+ E- $\to$ K0(S) K0(L) In The Total Energy Range 
%1.4-Gev To 2.18-Gev And Interpretation Of The K+ And K0
Phys.\ Lett.\ B {\bf 99}, 261 (1981);
%%CITATION = PHLTA,B99,261;%%
%
%\bibitem{Ivanov:82}
P.~M.~Ivanov {\it et al.}, preprint NOVO-82-50(1982);
%
%\bibitem{Dolinsky:vq}
S.~I.~Dolinsky {\it et al.},
%``Summary Of Experiments With The Neutral Detector At The E+ E- Storage 
%Ring Vepp-2m,''
Phys.\ Rept.\  {\bf 202}, 99 (1991).
%%CITATION = PRPLC,202,99;%%


%\cite{Akhmetshin:vz}
\bibitem{Akhmetshin:vz}
R.~R.~Akhmetshin {\it et al.},
%``Measurement Of Phi Meson Parameters With Cmd-2 Detector At Vepp-2m 
%Collider,''
Phys.\ Lett.\ B {\bf 364}, 199 (1995).
%%CITATION = PHLTA,B364,199;%%

\bibitem{Brodsky:1974vy}
S.~J.~Brodsky and G.~R.~Farrar,
%``Scaling Laws For Large Momentum Transfer Processes,''
Phys.\ Rev.\ D {\bf 11}, 1309 (1975).
%%CITATION = PHRVA,D11,1309;%%

%\cite{Fadin:zw}
\bibitem{Fadin:zw}
V.~S.~Fadin and V.~A.~Khoze,
%``Production Of A Pair Of T Anti-T Quarks Near Threshold,''
Sov.\ J.\ Nucl.\ Phys.\  {\bf 53}, 692 (1991)
[Yad.\ Fiz.\  {\bf 53}, 1118 (1991)].
%%CITATION = SJNCA,53,692;%%

%\cite{Hagiwara:pw}
\bibitem{Hagiwara:pw}
K.~Hagiwara {\it et al.}  [Particle Data Group Collaboration],
%``Review Of Particle Physics,''
Phys.\ Rev.\ D {\bf 66}, 010001 (2002).
%%CITATION = PHRVA,D66,010001;%%

\bibitem{Cheng:1999tr}
H.~Y.~Cheng and K.~C.~Yang,
%``updated analysis of a_1 and a_2 in hadronic two body decays''
Phys.\ Rev.\ D {\bf 59}, 092004 (1999) [arXiv:hep-ph/9811249].
%%CITATION = HEP-PH 0112245;%%

%\cite{Chua:2001xn}
\bibitem{Chua:2001xn}
C.~K.~Chua, W.~S.~Hou and S.~Y.~Tsai,
%``Prediction of three-body B0 $\to$ rho- p anti-n, pi- p anti-n decay  
%rates,''
Phys.\ Lett.\ B {\bf 528}, 233 (2002)
[arXiv:hep-ph/0108068].
%%CITATION = HEP-PH 0108068;%%

%\cite{Asner:1999kj}
\bibitem{Asner:1999kj}
D.~M.~Asner {\it et al.}  [CLEO Collaboration],
%``Hadronic structure in the decay tau- $\to$ nu/tau pi- pi0 pi0 and the  
%sign of the tau neutrino helicity,''
Phys.\ Rev.\ D {\bf 61}, 012002 (2000)
[arXiv:hep-ex/9902022].
%%CITATION = HEP-EX 9902022;%%



%\cite{Weinberg:1967kj}
\bibitem{Weinberg:1967kj}
S.~Weinberg,
%``Precise Relations Between The Spectra Of Vector And Axial Vector 
%Mesons,''
Phys.\ Rev.\ Lett.\  {\bf 18}, 507 (1967).
%%CITATION = PRLTA,18,507;%%

%\cite{Bjorken:kk}
\bibitem{Bjorken:kk}
J.~D.~Bjorken,
%``Topics In B Physics,''
Nucl.\ Phys.\ Proc.\ Suppl.\  {\bf 11}, 325 (1989).
%%CITATION = NUPHZ,11,325;%%

%\cite{Ligeti:2001dk}
\bibitem{Ligeti:2001dk}
Z.~Ligeti, M.~E.~Luke and M.~B.~Wise,
%``Comment on studying the corrections to factorization in B $\to$ D(*) X,''
Phys.\ Lett.\ B {\bf 507}, 142 (2001)
[arXiv:hep-ph/0103020].
%%CITATION = HEP-PH 0103020;%%

%\cite{Bauer:2002sh}
\bibitem{Bauer:2002sh}
C.~W.~Bauer, B.~Grinstein, D.~Pirjol and I.~W.~Stewart,
%``Testing factorization in B $\to$ D(*) X decays,''
arXiv:hep-ph/0208034.
%%CITATION = HEP-PH 0208034;%%



\end{thebibliography}
\end{document}